Politechnika Łódzka

Wydział Elektrotechniki, Elektroniki, Informatyki i Automatyki

Instytut Informatyki Stosowanej

PRACA DYPLOMOWA INŻYNIERSKA

# Design and implementation of a user interface for a multi-device spatially-aware mobile system

Przemysław Kucharski

181445

Opiekunowie pracy:

dr inż. Andrzej Romanowski

Paweł Woźniak, tekn.lic.

## *Abstract*

The aim of this thesis was the design and development of an interactive system enhancing collaborative sensemaking. The system design was based on related work research and preliminary user study. The system is based on multiple spatially-aware mobile devices. The spatial awareness is established with a motion tracking system. The information about position of tablets is received by a server. The server communicates with the devices and manages the content of each device. The implemented system supports managing the elements of information across the devices basing on their relative spatial arrangement. The evaluation of the system was done by a user study.

## *Streszczenie*

Tytuł pracy: Projekt i implementacja interfejsu użytkownika dla systemu opartego na wielu urządzeniach mobilnych wykorzystujących świadomość przestrzenną

Celem pracy był projekt i implementacja systemu wspomagającego proces nadawania znaczenia postrzeganym informacjom (ang. sensemaking) podczas pracy w grupie. Projekt systemu powstał w oparciu o dostępną literaturę i badaniu użytkownika przy użyciu prototypu o niskiej dokładności. Zaprojektowany układ jest oparty o urządzenia mobilne wykorzystujące świadomość przestrzenną, uzyskaną przy pomocy systemu do śledzenia ruchu. Informacje na temat pozycji są dostarczane do serwera, który komunikuje się z urządzeniami i zarządza wyświetlaną treścią. Ewaluacja systemu została wykonana w formie badania użytkowników.

# Table of contents

## *Acknowledgements*

I would like to thank my supervisors, dr inż. Andrzej Romanowski and Paweł Woźniak, tekn.lic., for their support and valuable advice. I also thank the Team of t2i Interaction Laboratory for the great time spent working with them in Gothenburg.

# 1. Introduction

In a world of fast technological development, mobile devices are becoming more portable, powerful and affordable. A proliferation of mobile devices can be observed both at work and in our homes. Consequently, the range of usage scenarios for mobile devices is constantly extending. We use our phones and tablets almost everywhere and at any time. It comes with no surprise if we meet friends or come to a meeting at work and each person in the room carries a mobile device. It is also not uncommon to spot a person using two devices simultaneously - with a messaging app on a smartphone and a news feed on a tablet.

The presence of technology almost everywhere affects the way people explore data. We are constantly connected and often we may not even notice using technical solutions to perform everyday tasks. The pace of life nowadays makes it important to be able to cope with big amounts of information efficiently at all times. This means that technology should, to a great extent, support the process of data exploration and sensemaking. This ubiquity of mobile devices and the variety of their possible combinations pose new challenges to the field of Human-Computer Interaction (HCI). This work explores possible ways we will use our phones and tablets in the future.

This work is primarily inspired by the continuous mobile technology development broadening the range of sensors, making it possible for devices to interact with the environment in a new manner. It is critical now to examine in depth how this possibilities should be used for user benefit.

The solution presented in this work was implemented as a part of a research internship in t2i Interaction Laboratory, Chalmers University of Technology, Gothenburg, Sweden. The idea of the project evolved from past research on multi-device ecologies performed by the team of the lab. The author of this thesis completed a research secondment in the laboratory, directly engaging in the day-to-day operations of the unit.

The detailed goal of this work is presented in section Design. The motivation for this work is given after Related Work.

## 2. Related Work

The topic of data exploration using mobile devices and interactive tabletops has been a subject of past research, and many examples of related work can be found. In this section, several publications of relevance for this field are considered.

### 2.1. Data exploration

Fjeld et al. [9] raised the issue of „big data" present in public space and the emerging issue of information becoming inaccesible to most people. They discussed the design considerations for ad-hoc data exploration and proposed tangible tabletops as one of possible solutions, with an insight of future technology development for interactive environments.

Weise et al. [33] investigated how the issue of data administration, exploration and analysis should be addressed in the age of ubiquitous computing. They suggested connecting it to the local infrastructure, both in terms of human awareness and environmental sensing.

### 2.2. Multi-device systems

As stated before, situations when users have several mobile devices to work at the same time can often be encountered. Currently available applications, however, do not provide solutions by which users can benefit from the interactions between the devices they use, i.e. cross-device interactions.

The subject of multi-device environments was considered for a long time. Bilezikjian et al. [4] explored how the interactions with handheld devices may look like, although they were not commercially available. Blackwell et al. [5] investigated the issue of tangibility in the context of mobile devices. Cauchard et al. [7] discussed the concerns and opportunities for visual aspect of mobile multi-display environments.

In Conductor [14], Hamilton and Wigdor presented a framework for examining the scenarios of cross-device interactions. They provide functionalities to split the aspects of performed task between several mobile devices. They also elaborated possible usage scenarios for the system, demonstrating their way of understanding cross-device interactions in task-specific domain. The study performed using the developed multi-device system demonstrated that using several connected devices is highly useful to perform certain tasks. Most of the participants made use of multiple devices and of the functionalities enabling them to easily transfer information across devices.

The Pass-them-around [19,20] system developed by Lucero et al. showed that providing people with the functionality of sharing content between their personal devices may lead to enrichment of interactions between users.

Cassens et al. [6] proposed a taxonomy for the term cross-device interactions. They introduced the dimensions of ownership, distance and access to classify those interactions. They discussed this classification in the context of the work published in this field. Finally, they proposed a definition: *"Cross-device interaction (XDI) is the type of interaction, where human users interact with multiple separate input and output devices, where input devices will be used to manipulate content on output devices within a perceived interaction space with immediate and explicit feedback."*

In general, the past research in the field of multiple device ecologies demonstrates that once cross-device interactions are implemented and enabled, users adapt to the new environment and benefit from these interactions. This proves, as it was raised in many of the papers, that further exploration of this field is necessary and promises a valuable contribution to our future life.

### 2.3. Multiple Surface Environments

In [1,2], Ballagas et al. investigated the way people may use their personal mobile devices to interact with surfaces such as big displays present in public spaces. This work focused on using personal devices as medium of interaction rather than interacting with the public screen in a physical way, a situation we can meet often in everyday life.

Ballendat et al. [3,11] also focused on interacting with big vertical surfaces, but investigated in more detail how those interactions can be designed taking into considerations the proximity of the user and the screen. Their approach based on transferring the theory of proxemics, developed by Hall [13] as a way of understanding the relative spatial relations in human-to-human interactions, to digital domain.

Marshall et al. [22] conducted a long-term in-wild study to examine the ways groups of people collaborate over an interactive tabletop. Their findings showed different approaches and scenarios of using their application called Tourist Planner. The important conclusion of this work is the spotted difference between how people use tabletops during a laboratory study and in the wild.

Wallace et al. [32] investigated the field of using mobile devices and tabletops. Their work consisted of comparison between three distinctive scenarios of used displays:

mobile devices, a tabletop, and both mobiles and a tabletop. The study they conducted demonstrated the relations between the task completion performance and teamwork in collaborative scenarios. One of the most significant findings of this work is that the presence of shared devices increase the performance compared to the scenario where users collaborate using devices as their personal ones.

This part of the related work focused mainly on the design of interactions for ubiquitous computing ecologies, and shows that proper investigation of the field of collocated collaboration over interactive displays may influence not only the way people use technology, but also how they interact with other users.

## 2.4. Spatially aware systems

Space and awareness of space is of great importance for human perception of surrounding world. Hall developed a theory describing the relation between spatial arrangement of people and their social behaviours and emotions. Chen and Kotz [8] included spatial awareness as an important aspect of context-aware computing. Some past research show that space may also have a vital impact on understanding and learning processes [26]. There was also some investigation concerning how transferring the perception of space to technology may help users interact with data.

In MochaTop [34], Woźniak et al. made a step forwards to understand how spatial combinations of two devices can be used for data exploration. They focused on a single user scenario for ad-hoc interactions, basing on an assumption that a couple of a smartphone and a tablet is more and more often carried by users. They designed and implemented several spatial-based interactions for exploring complex structures of data.

Their study showed that users find it useful to make space one of the input sources in data manipulation.

Spindler [28–30] proposed a way of extending the interaction space into the third dimension. In this work, users are in disposition of several spatially aware tangible displays to interact with virtual objects present on a central tabletop. This work demonstrated the potential of employing explorable 3D space for visualisations. By direct translation of spatial position to visualisation input, he created a tool for intuitive exploration of several types of complex datasets.

HuddleLamp [24] provides a solution for creating ad-hoc multi-device ecologies on a tabletop based on video processing. Rädle et al. propose spatially-aware multiple device systems as an alternative for interactive tabletops. An interesting approach in this work is tracking not only the devices, but also the motion of users' hands to enrich interactions.

The field of extending the interaction space of one mobile device was also investigated. AD-binning [15], which is an abbreviation from Around-Device Binning, is a mobile user interface that allows users organise data spatially outside of the device space. The concept consists of creating virtual zones (bins) around the device. The system tracks relative position of user's finger and the device and thus makes it possible to put pieces of information from the screen into these zones. The motivation of designing such solutions is, as stated by the authors, extending the interation space due to insufficiency of that space on the screen of a mobile device.

The premise that the design of user interfaces for multiple mobile devices should be based on spatial awareness of those devices was discussed by Rädle et al. [25]. They divided cross-device interactions into three groups: spatially agnostic, based on synchronous gestures and spatially aware. Spatially agnostic interactions may be menu-based with some kinds of user-perceivable device identification. Interactions based on synchronous gestures are triggered by simultaneous interaction with two of more devices. Eventually, spatially aware interactions are based on relative arrangement of devices. The study conducted to compare them demonstrated that in most cases, spatially-aware interactions between devices are expected by the users. However, similarly to the Ballendat's conclusion about proxemics, Rädle et al. clearly emphasise that the way in which spatially aware interactions should be designed is not yet fully explored and requires further investigation.

## 3. Motivation

Changing world poses new challenges to researchers in the field of HCI. The increasing importance of ad-hoc data exploration motivates me to address those challenges in this particular direction.

Conclusions drawn from past reseach show that cross-device interactions require an extensive investigation to make the effects of design and development most efficient and usable.

I joined a team of researchers with a considerable expertise in investigating spatial cross-device interactions. Starting from the experience of recent projects realised in the laboratory, through analysis of achievement of fellow teams in this field, a decision was made to develop and investigate a new generation of multi-device interactive system.

The synthesis of past research, having in mind spatiality of interaction and number of users for which the system is designed, clearly indicates the path of research which should be taken to make the base of knowledge about mobile device interactions more consistent [Table 1]. Several non spatially aware systems were developed and evaluated, both in individual and collaborative work. However, the literature lacks sufficient investigation of multi-user systems that use sptaial awareness as an aspect of interaction.

*Table 1: Comparison of already developed systems*

|  | Non spatially-aware | Spatially-aware |
|---|---|---|
| Single-user | Conductor | HuddleLamp |
| Multi-user | Pass-them-around | ? |

Designing and developing systems that support data exploration is very important for the range of problems where human expertise, and often intuition, is required to find the proper solution. This is closely related to the proces of sensemaking [10,27], i.e. giving meaning to the observed data. Aiming at the sensemaking proces to be more efficient may be meaningful for processes such as thematic analysis and decision making. A part of sensemaking is creating an associative network in the human mind. This is why spatial organisation of information is of high relevance for this proces, which was studied by Keel [17], who developed a system for inferring the relations and relevance of information from its spatial arrangement and collaborative use.

These insights were fundamental for the direction of research project conducted previously in the team I joined. The developed systems: DynamicDuo [23], MochaTop [34] and Thaddeus [35], were aiming at investigation of how spatial interactions may help people understand the information presented to them.

The previous systems were based on two spatially-aware mobile devices, defined as the central and satellite device. The position of the satellite determined the element explored on the central device – the details were diplayed on the screen of satellite device. The results achieved with these systems were highly motivating to continue research based on this experience

Another motivation for this kind of projects is the perspective of developing publicly available systems engaging spatial awareness. The technology of sensing the surroundings of a mobile device was developed by Elliptic Labs [36]. It is based on ultrasound sensing of the space around the device. The whole sensor system is compact enough to include it in the device's body. The software developed by this company recognises the movements around the device. It was confirmed that this kind of technology will appear in commercially available devices in several years.

# 4. Design

## 4.1. Design goal

The goal of the design is to develop a system enabling collaborative work [12] with multiple mobile devices, such as tablets, on a horizontal surface. As stated before, this system should be in the form of a user interface distributed in terms of display and input spaces. The novelty in this system is the implementation of spatial awareness of devices, which opens new directions in the design of the interface.

## 4.2. The design space

Possibly the most important of the aforementioned directions given by spatial awareness is the possibility of including the whole surface of tabletop into the interaction space, not only the space of the devices' displays, as in traditional interfaces. Therefore we think of the interface of the entire space, for example a table on which the devices are placed, and devices may be considered as "islands" on which the proper part of the interface is currently displayed, basing on the position of the device. The interface, thus, should be designed in a way that the user can easily imagine the invisible part of interaction space, so that the perception of information displayed on the devices is easier [Figure 1].

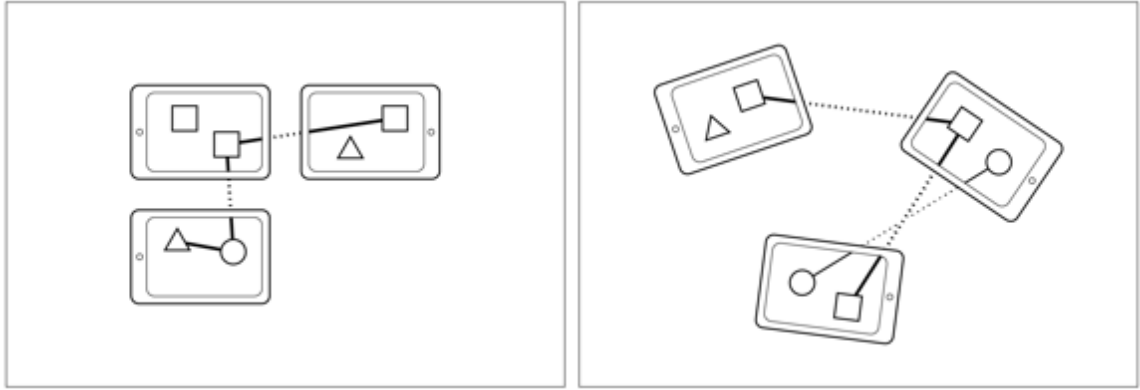

*Figure 1 Using the entire surface as design space*

The dataset considered in design of this system is a set of crisp pieces of information, such as text notes, photos, images, video clips etc. It is assumed that elements of the set can be automatically analysed in order to create metadata, which makes it possible to define initial relation between them. These relations may have spatiotemporal forms or be based on common references to people, things etc.

### 4.3.Interaction design

In order to determine the desired way of interaction with the system, an initial user study was conducted. It revealed that users tend to organize pieces of information spatially according to the criteria mentioned above Figure 2. Also, users claimed that if they were given a way to show these relations in other manner, they could use the space in other way, which would make the process of understanding easier.

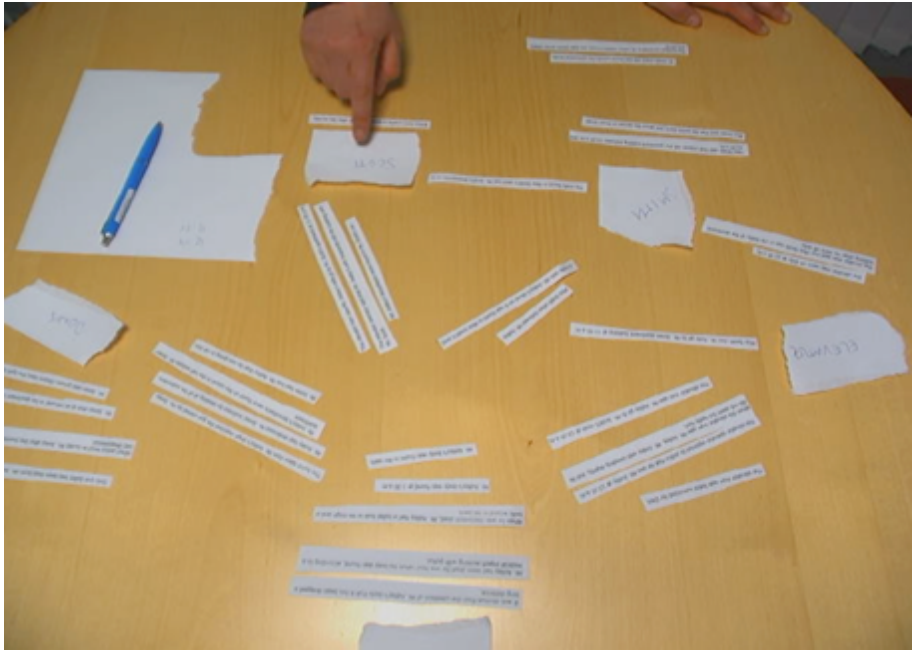

*Figure 2 Spatial organisation of information (Preliminary user study)*

### 4.4. Final design

After conducting user study in initial phase and another study in early phase of system implementation, the conclusion was that the system should enable both manipulating the data and showing relations between them across the devices.

User should be able to easily move elements from one device to another, basing on the relative spatial position, and to see the relations between elements in an understandable form. Passing an object between two devices is realized as a quick swipe of the object in the desired direction [Figure 4].

Relations based on common references should be realised as distinction of all related elements on all devices and are triggered by a click on the element, while temporal relations should connect the elements ordered timewise through the whole interaction space, i.e. surface of the table [Figure 3]. The latter interaction is triggered by a long click on the element.

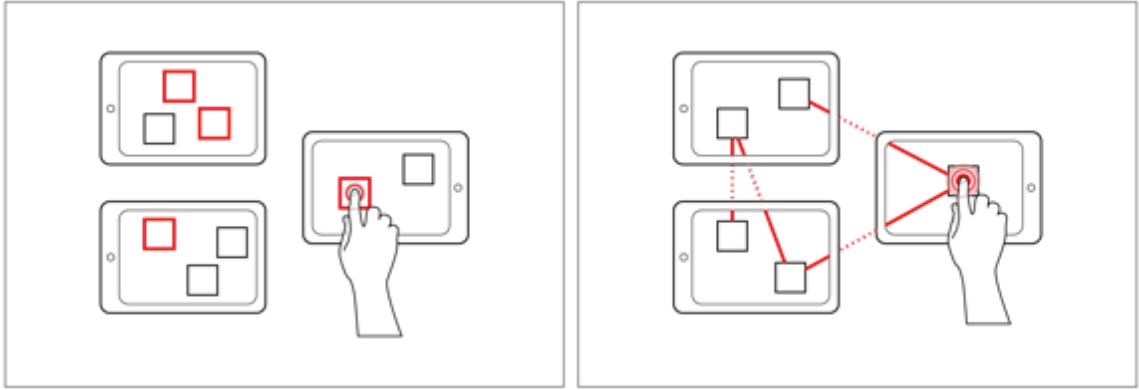

*Figure 3 Higlighting related elements and displaying a timeline*

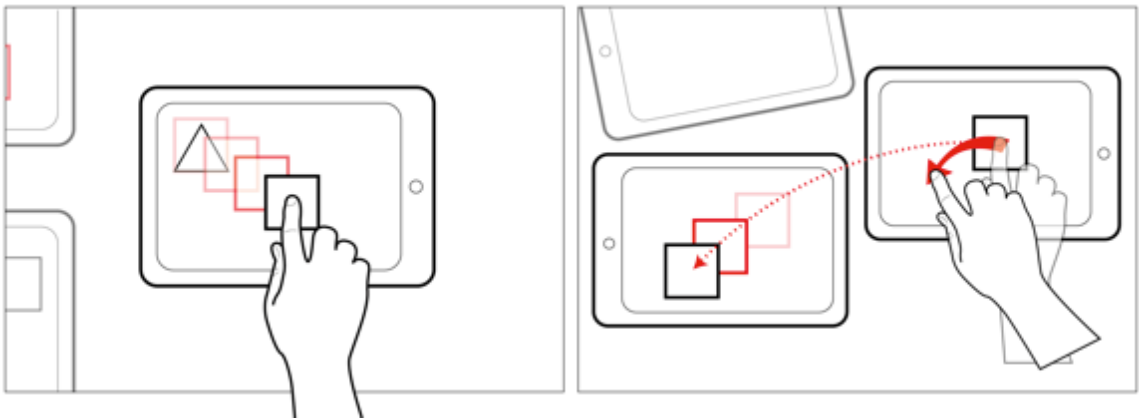

*Figure 4 Moving the elements on and between the devices*

# 5. Implementation

## 5.1. Structure of the system

The most important part of the system is the set of mobile devices placed on the table. An external motion tracking system provides position data in real time by its built-in server. Due to the complexity of this setup, I decided that it is necessary to introduce my own server running on a separate machine, which will coordinate the whole system.

Therefore the setup on which the system was implemented consists of three Android tablets, a laptop with motion tracking software and laptop with my server. All devices were connected to one local area wireless network. The hardware structure of the system is depicted in Figure 5. The main element of the system is `RampartsServer`, which receives position data from the motion tracking system and communicates bidirectionally with the devices.

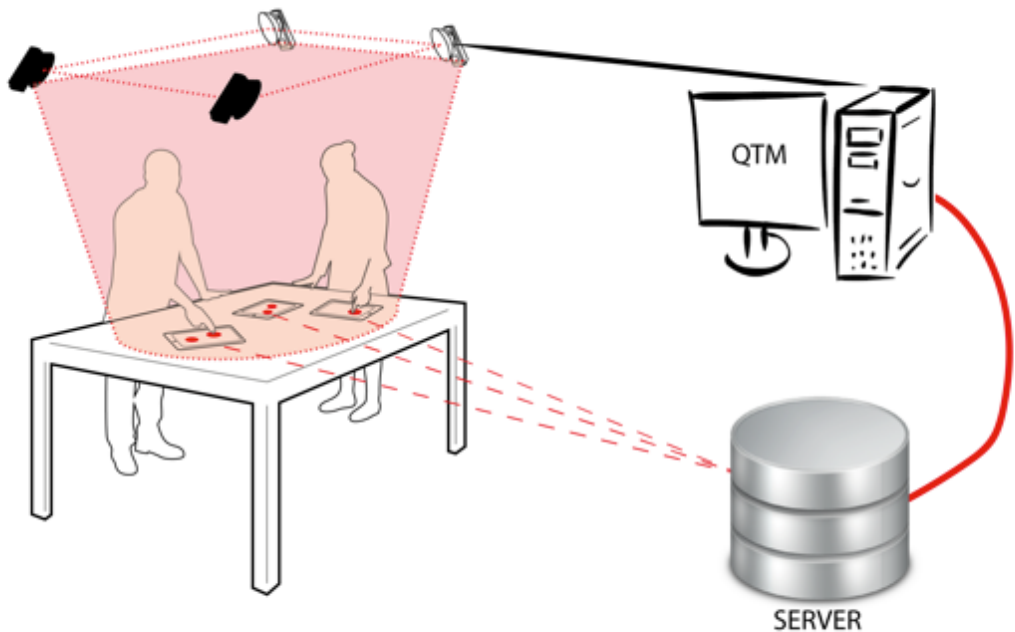

*Figure 5 Hardware structure of the system*

## 5.2. Motion tracking system

For implementation of spatial awareness in the system, I used motion tracking hardware and software provided by Qualisys AB [37]. It is a commercial product which is capable of giving precise and fast position data. The available setup consisted of 8 Oqus cameras, set of passive markers and Qualisys Track Manager (QTM) as measurement software.

### 5.2.1. Oqus cameras

The cameras used in the motion tracking system are high resolution devices that use infrared light sensing [Figure 6]. The model installed in the available setup support resolution up to 4 megapixels with framerate 180 fps. The cameras are daisy chained; one end is plugged into a computer via Ethernet cable. The main feature of the cameras is the ability to capture the position of markers and transfer the image to the Qualisys software.

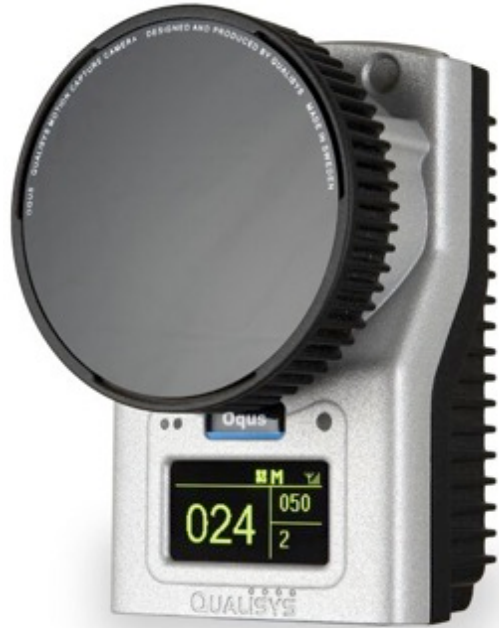

*Figure 6 Oqus series camera*

### 5.2.2. Markers

For tracking any object with the Qualisys system, it is necessary to mount retro-reflective markers on its body [Figure 7]. The system tracks these markers and all further processing is based on their position, therefore it is crucial to determine the best placement of markers on the body in terms of what information is needed. Also, it should be taken into account that markers should be visible for the cameras for as much time as possible during motion tracking.

In the developed system, passive reflective markers of 7mm diameter were used. They are flat base spheres covered with retro-reflective tape. They were mounted on the edges of devices with double-side tape.

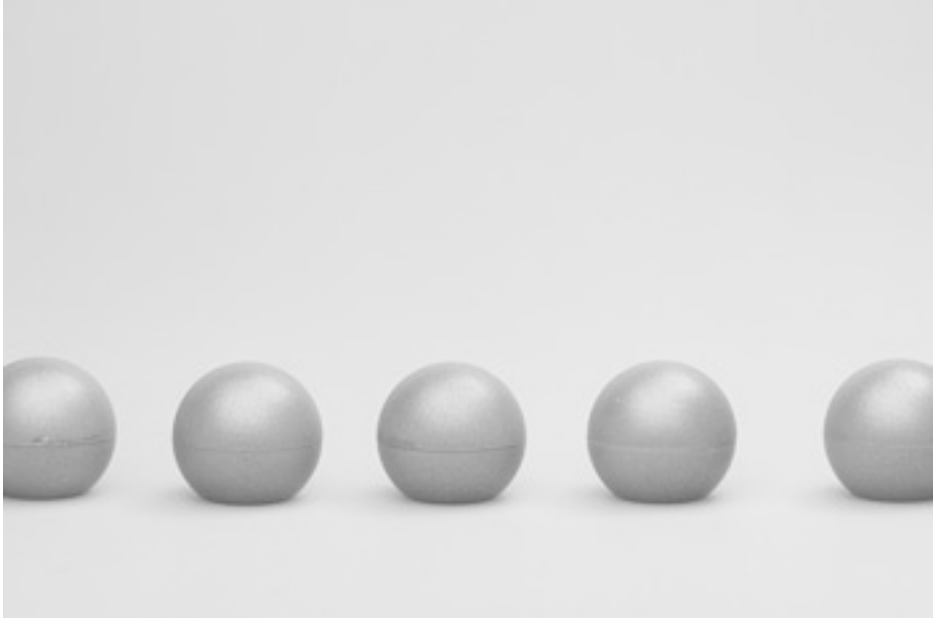
*Figure 7 Retro-reflective markers*

### 5.2.3. Qualisys Track Manager

QTM [Figure 8] is the main software that handles the process of motion capture with camera system. It collects information from all cameras connected via an Ethernet interface and offers many functionalities of data analysis and output.

#### Main features

QTM features 2D, 3D and 6DOF data tracking with latency down to 4 ms. It supports both passive and active markers. Also, automatic marker identification and marker masking is possible with the software. QTM automatically detects the number of cameras. Users can see the images from all cameras and mask the regions with unwanted noise. For 3D and 6 Degrees Of Freedom (6DOF) tracking, the system needs to be calibrated. To find the position of a marker in 3D space, the software analyses the input from all cameras. Calibration is the process of identifying the space within the camera

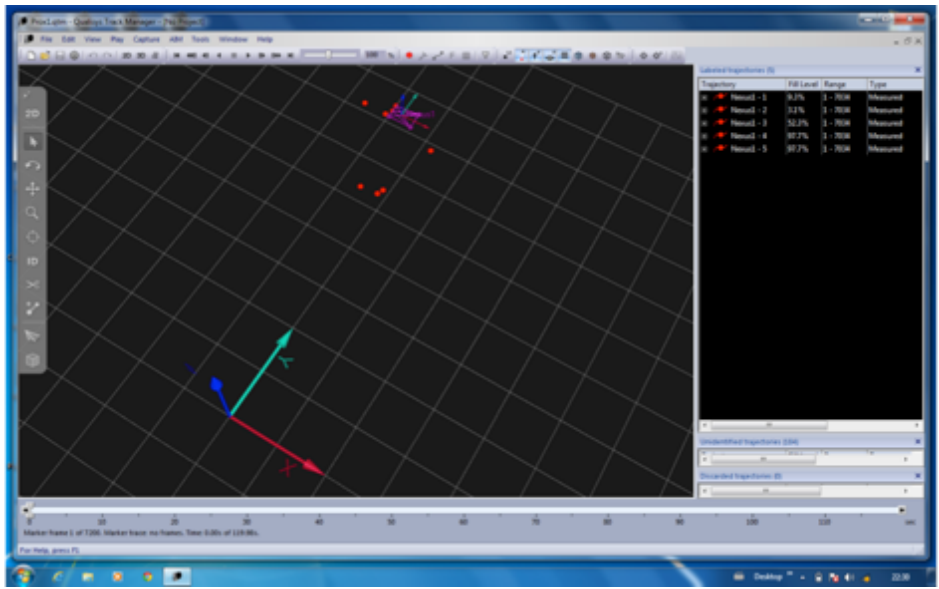
*Figure 8 Qualisys Track Manager interface*

system. It is performed using two tools: an L-frame, which contains four markers identifying the axes - the L-frame should be located in the origin of desired coordinate system. The other tool is the calibration wand, which consists of two markers at the edges of a T-shaped tool [Figure 9]. The dimensions of both the L-frame and the wand should be entered to the system. The process itself involves moving the wand inside the space to be calibrated, with the idea that the markers should appear in as many points as possible. The way of moving the wand can be called 3D painting – user should try to "paint" each point in space with the markers on the wand. Knowing the distance between wand markers, the system is able to learn how to transform 2D data from cameras into 3D image.

Qualisys Track Manager supports recording and real-time output. Captured markers may be automatically identified if such a feature was configured in project options. An important functionality of the software is ability to define a rigid body from several markers. It makes it possible to export 6DOF data. From the point of view of reliability of tracking, defining rigid bodies supports tracking even if not all the markers are visible at some point in time. This is extremely useful for situations when some of the mrkers may be obscured e.g. by user's hand.

Furthermore, QTM supports video recording and data export in several formats, such as .tsv, .c3d, and .mat files.

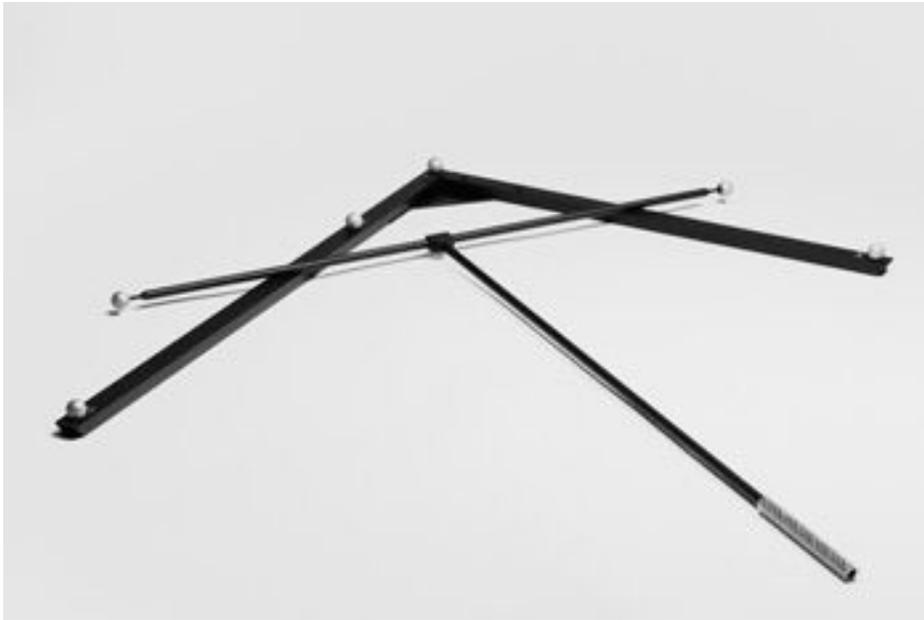
*Figure 9 Qualisys calibration kit*

### 5.2.4. QTM Real-Time Server

Qualisys Track Manager offers the functionality of real-time data retrieival via TCP/IP or UDP/IP connection. To use that, one should connect to the proper port on QTM-running machine. The parameters of data streaming may be configured using this connection. Server may be configured either in streaming or polling mode. For each type of data QTM can provide, the structure of data packet sent in the network stream is specified.

### 5.2.5. Available setup

The setup to which the author of this work had access during the implementation consisted of 8 Oqus 500+ cameras located as shown in Figure 10. A computer with both anEthernet port and wireless network card was used to run Qualisys Track Manager.

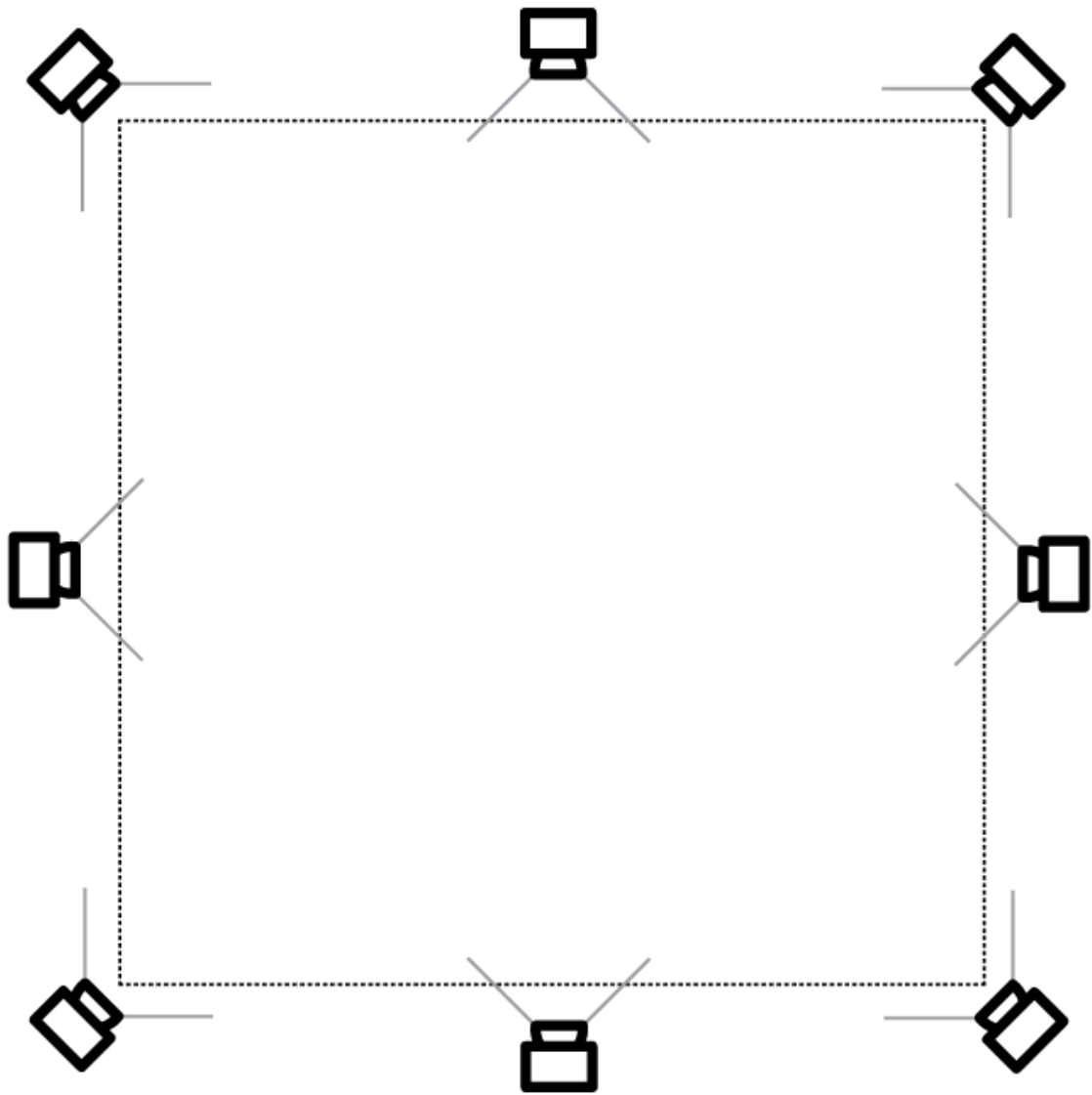

*Figure 10 Available setup*

## 5.3. Server

### 5.3.1. Development

The server was developed using Eclipse IDE for Java Developers [38][Figure 11]. The server is written in Java SE 8. Gitorius was used for version control. Apart from Java native libraries, I used Jama [39], an open source mathematical library for matrix operations.

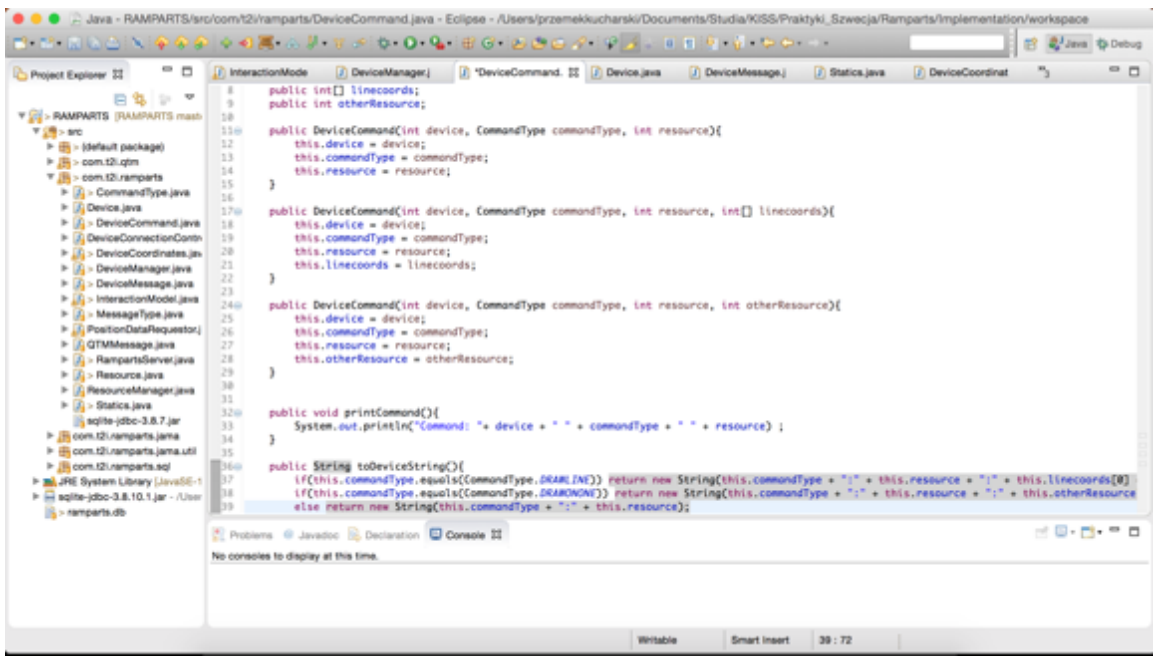

*Figure 11 Eclipse IDE*

### 5.3.2. Main functionalities

As mentioned, the main purpose of the server is to coordinate the behaviour of tablets. To do this, the server has to establish bidirectional communication with them, and get information about spatial arrangement of devices.

This purpose is reflected in the main class of the server. This class is running two main threads - one communicating with QTM, and the other managing the behaviour of devices. Selected classes of the server and their structure are depicted in Figure 12.

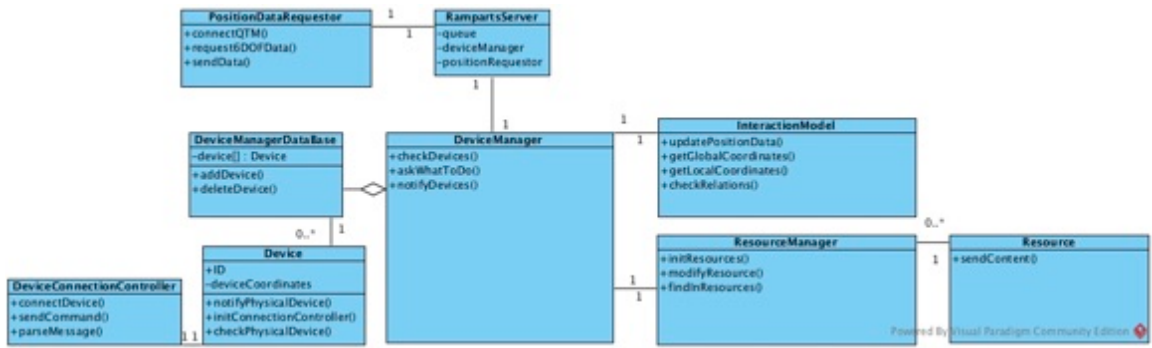

*Figure 12 Selected classes of server*

### 5.3.3. Requesting position data

The class `PositionDataRequestor` connects to machine with QTM, to the port specified in QTM enabling to get data from RT Server. Because each tablet is defined as rigid body in QTM, basing on markers mounted as shown in Figure 13, it is possible to receive 6DOF data. The server is set in requesting mode, i.e. it sends one frame of data each time it is requested from my server. The data are then parsed so that we receive a frame of devices. For each device, the coordinates X, Y and Z of its geometrical centre are given, as well as three Euler angles. These data are then transformed into two-dimensional array of coordinates and put in a form of `QTMMessage` in a `SynchronousQueue`, from which the thread managing the devices takes them.

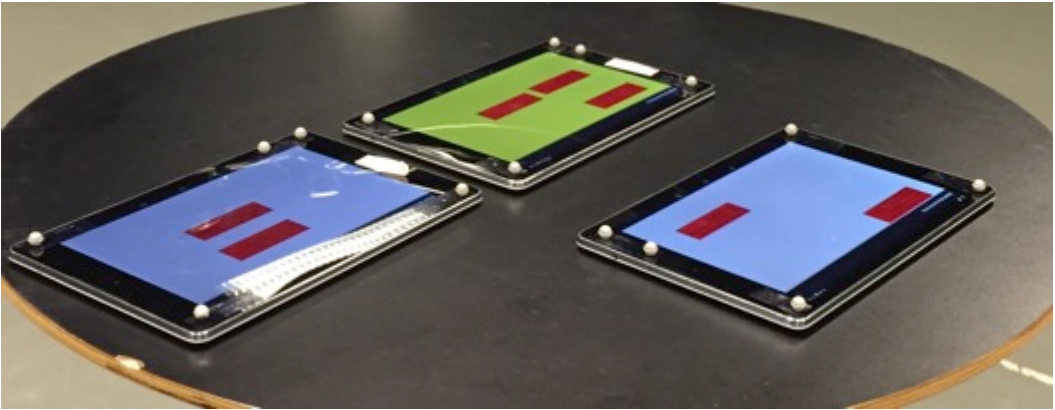

*Figure 13 Devices with markers*

### 5.3.4. Managing the devices

Each tablet is assigned to an object of `Device` class in `DeviceManagerDataBase`, an internal class of `DeviceManager`. It stores the current position of the device and is responsible for bidirectional communication with tablets, using native Java Socket. `DeviceManager` pulls an object of `QTMMessage` and updates the coordinates of each device. It also gathers the messages from all tablets, sends it to an object of `InteractionModel`, class responsible for making decisions about what should happen on each device, and communicates the result of `InteractionModel` work to tablets.

### 5.3.5. Database connection

For further studies using this system, I implemented the server in a way it can communicate with an SQLite database. The created database can store all information about resources and devices. It is initialised at the start of the system. Several classes in the system, among others `InteractionModel` and `Device`, invoke the instance of a singleton class with methods containing SQL statements to communicate with the database.

## 5.4. Mobile application

### 5.4.1. Implementation platform

The implementation platform chosen for this project was Android [40]. It supports programming in Java language. It provides a framework to properly configure applications for different devices. It has a rich base of resources and libraries that make the development proces more efficient. As it is an open-source project, building custom libraries and components is also easy. The project was developed for Android 5.0 Lollipop.

### 5.4.2. Development

The mobile application was developed using Android Studio [41,42], the official IDE for Android application development. It supports among others Gradle-based build system and, which was important in preparing the system for user study, as it enables generating multiple .apk files. I also used Gitorius for version control.

### 5.4.3. Application features

The application running on mobile devices is designed mainly to communicate the interactions to the server and to respond to commands in a proper manner.

Each device is given an unique ID, reflecting the ID of the device object on the server. After launching, the application connects as a client to a server socket on the port determined by ID.

The application creates the interface [Figure 14] on the screen. In the software version prepared for user study the interface consists of two main components: an `ImageView` in the background, making it possible to draw lines on the screen, and a set of `TextViews`, simulating post-it notes. The application creates the number of post-it notes specified, and makes several of them visible, basing on the device ID number. At the beginning, all parameters are set, such as initial color, size, position on screen, and features like clickability and longclickability.

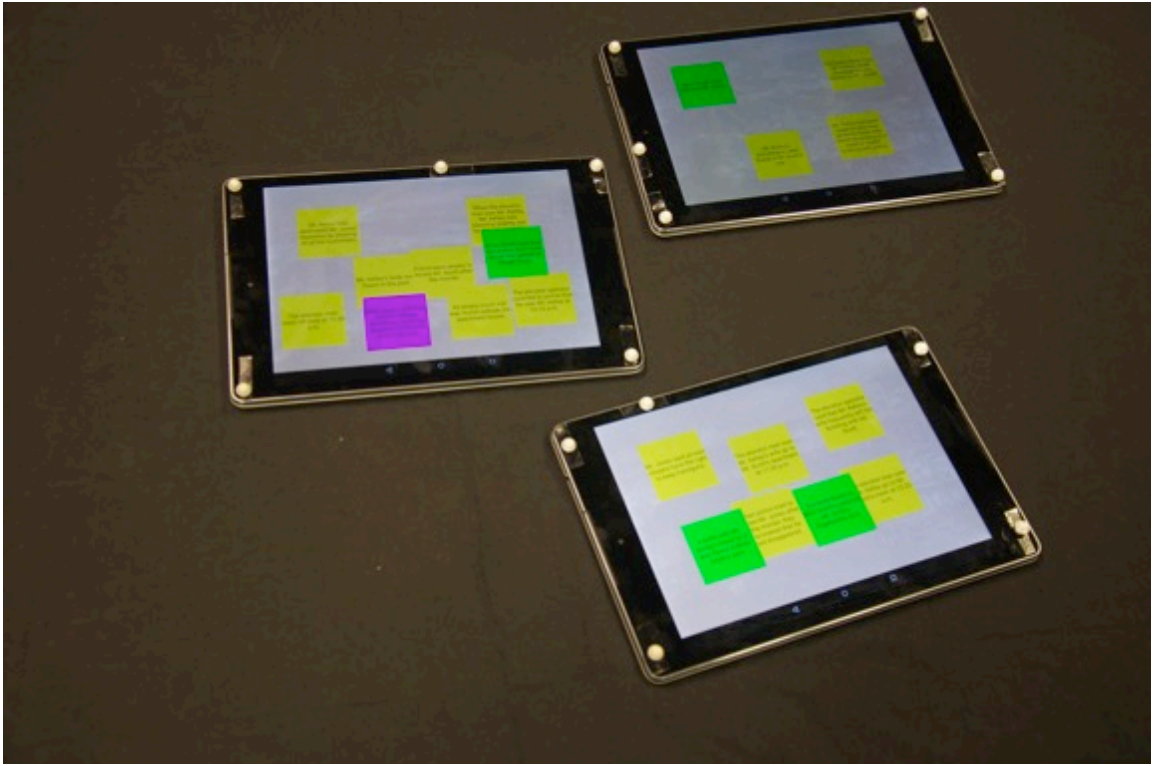

*Figure 14 Interface of mobile application*

Each post-it note is assigned three main event listeners important for the system. Methods `onClick` and `onLongClick` contain negation of corresponding global flag and assignment of resource id to corresponding global variable. This serves for determining the proper message sent to the server.

The `onTouchListener` in this application handles moving the resource, updating its position on the server side, and throwing it to another device. The algorithm for moving the resource was developed in a way that for movement in each direction, the method check if user touches the object close to the corresponding edge and then moves the resource on screen. This way of implementation easily avoids unwanted moving and hopping of the view.

In parallel, if a move gesture happens inside the object, the velocity of this movement is calculated. In case it exceeds certain value, the flag that a resource is being thrown is set and the ID of the resource and X, Y values of velocity are saved to send them to the server. If throwing does not happen, the application knows that the resource is just being moved and sets the corresponding flag. Also, new coordinates of the centre of the post-it note are sent to the server.

The message to server is determined before each sending it through a socket basing on which flags are currently set.

The thread communicating with server also receives commands. Each new send of commands invoke the method, which calls a handler on the UI and modifies what is visible on the screen basing on what the server ordered.

### 5.5. Communication between tablet and server

An important issue in the operation of the system is the communication between devices and server. The structure of communication in the system is shown in Figure 15. Each device is given a unique identification number, and connects to a separate port on the server machine. Once the connection is established, the device starts sending messages about its current status, and the server starts sending commands to device.

### 5.5.1. Device messages

Server receives information in case:

- resource is clicked
- resource is long clicked
- resource is thrown away
- resource is moved

A class for handling these messages has been implemented both on tablet and server side. It contains the type of message, which is one of the aforementioned four cases or null, which indicates that no interaction takes place. It contains also the ID of activeresource and information specific for message type, e.g. coordinates on the device screen for the moving message.

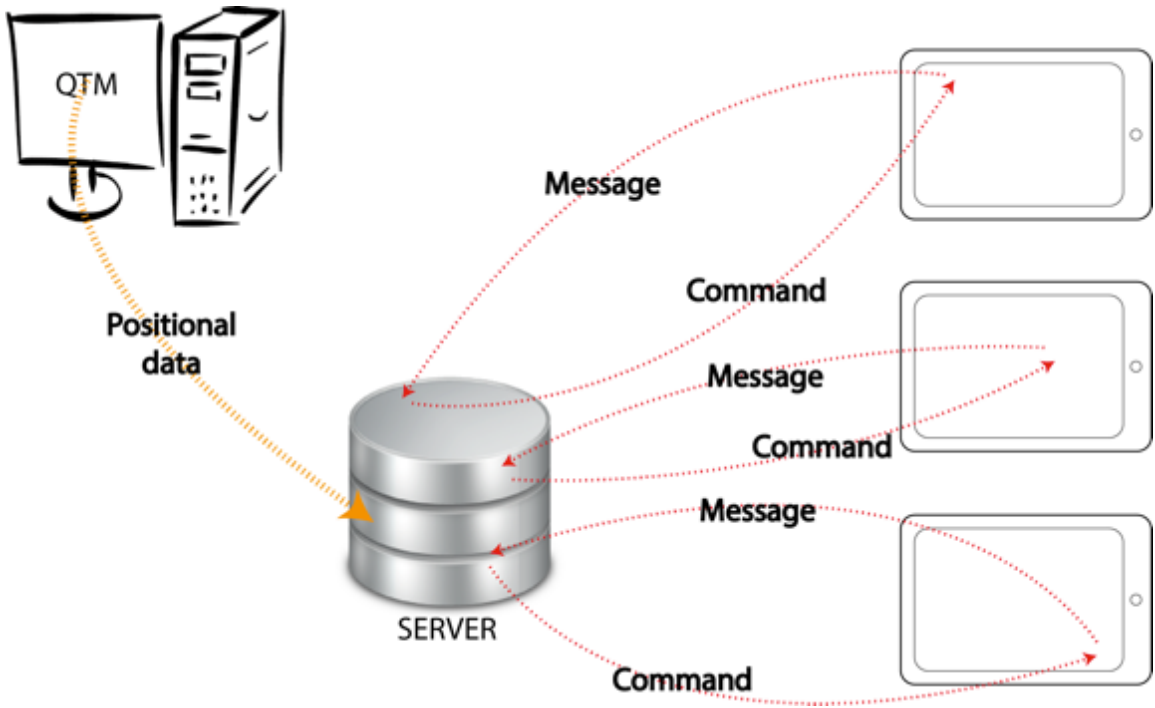

*Figure 15 Communication structure of the system*

### 5.5.2. Server commands

The structure of server command is similar to device message. Server may tell the device to:

- highlight a resource
- draw a line between resources
- show resource

In each case, the message contains the type of command and the ID of the resource on which it should be performed. In case of drawing lines, there are two types of commands: when both resources are on the same screen and when they are on different devices. For the fist case, IDs of both resources are sent, and in the other, ID of resource and coordinated of the second point are contained in the command.

## 5.6. Modelling the interaction

The most important issue in this system is making the decision about what each device should display, basing on the messages received from all devices. To do this, a class called `InteractionModel` has been implemented.

### 5.6.1. Data stored in `InteractionModel`

`InteractionModel` stores the information about position of each device, updated from each object of `Device` class. It stores also information about the relations between resources, and the position of each resource in global coordinate system.

### 5.6.2. Positioning resources

To handle the interactions properly, it is necessary to determine the position of each resource not only on the screen of device, but also in global coordinate system [Figure 16].

`InteractionModel` stores information on which device the resource is currently visible. On each movement of resource on the screen, the device sends message about new position in local coordinate system. Having this and the information about the position of device, the position of resource in global coordinate system can easily be calculated [18,21,43].

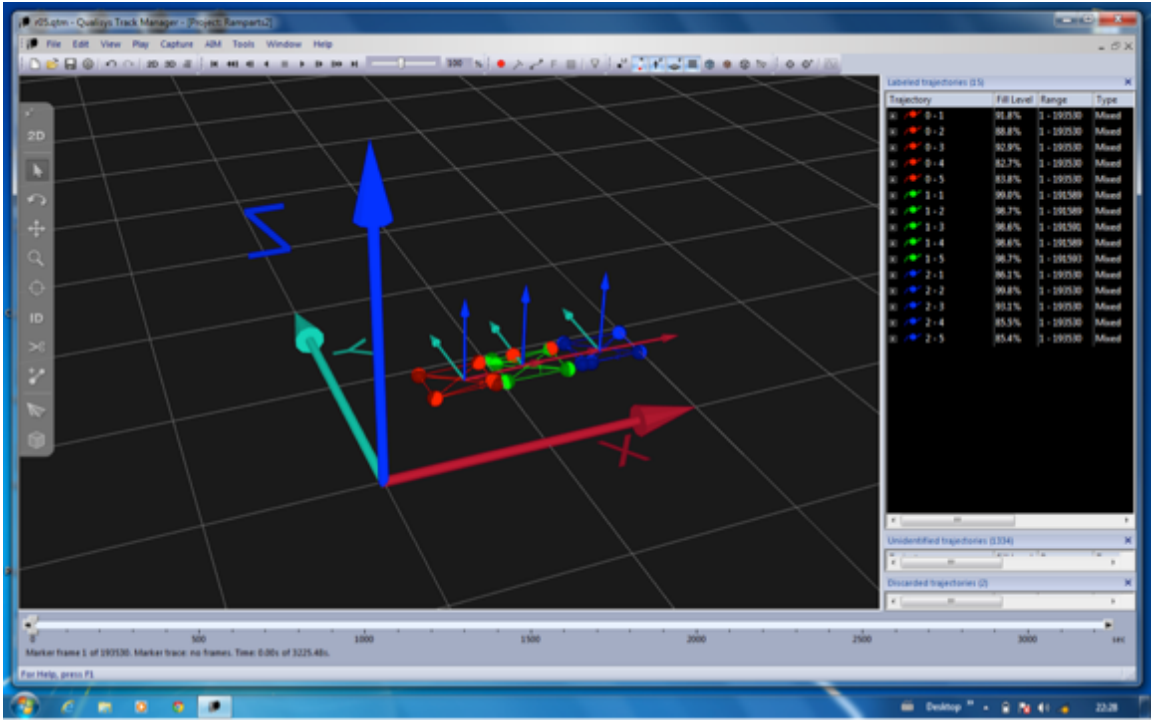

*Figure 16 Global and local coordinate system visible in QTM*

Having three Euler angles, properly configured on the motion tracking system side [44], we can perform rotations in the proper order:

$$\begin{pmatrix} X'' \\ Y'' \\ Z'' \end{pmatrix} = \begin{pmatrix} 1 & 0 & 0 \\ 0 & \cos\gamma & -\sin\gamma \\ 0 & \sin\gamma & \cos\gamma \end{pmatrix} \begin{pmatrix} x \\ y \\ z \end{pmatrix} = [R_1] \begin{pmatrix} x \\ y \\ z \end{pmatrix} \quad (5.1)$$

$$\begin{pmatrix} X' \\ Y' \\ Z' \end{pmatrix} = \begin{pmatrix} \cos\beta & 0 & \sin\beta \\ 0 & 1 & 0 \\ -\sin\beta & 0 & \cos\beta \end{pmatrix} \begin{pmatrix} X'' \\ Y'' \\ Z'' \end{pmatrix} = [R_2][R_1] \begin{pmatrix} x \\ y \\ z \end{pmatrix} \quad (5.2)$$

$$\begin{pmatrix} X \\ Y \\ Z \end{pmatrix} = \begin{pmatrix} \cos\alpha & -\sin\alpha & 0 \\ \sin\alpha & \cos\alpha & 0 \\ 0 & 0 & 1 \end{pmatrix} \begin{pmatrix} X' \\ Y' \\ Z' \end{pmatrix} = [R_3][R_2][R_1] \begin{pmatrix} x \\ y \\ z \end{pmatrix} \quad (5.3)$$

These three rotations can be combined into one rotation matrix:

$[R] = [R_1][R_2][R_3] ==$

$$\begin{bmatrix} \cos\alpha\cos\beta & \cos\alpha\sin\beta\sin\gamma - \sin\alpha\cos\gamma & \cos\alpha\sin\beta\cos\gamma + \sin\alpha\sin\gamma \\ \sin\alpha\cos\beta & \sin\alpha\sin\beta\sin\gamma + \cos\alpha\cos\gamma & \sin\alpha\sin\beta\cos\gamma - \cos\alpha\sin\gamma \\ -\sin\beta & \cos\beta\sin\gamma & \cos\beta\cos\gamma \end{bmatrix} \quad (5.4)$$

The rotation is followed by translation by the vector containing the position of device; we can combine all these calculations into one homogenous transformation matrix $T$:

$[T] =$

$$\begin{bmatrix} \cos\alpha\cos\beta & \cos\alpha\sin\beta\sin\gamma - \sin\alpha\cos\gamma & \cos\alpha\sin\beta\cos\gamma + \sin\alpha\sin\gamma & x_t \\ \sin\alpha\cos\beta & \sin\alpha\sin\beta\sin\gamma + \cos\alpha\cos\gamma & \sin\alpha\sin\beta\cos\gamma - \cos\alpha\sin\gamma & y_t \\ -\sin\beta & \cos\beta\sin\gamma & \cos\beta\cos\gamma & z_t \\ 0 & 0 & 0 & 1 \end{bmatrix} \quad (5.5$$

$)$

Finally, we obtain the formula to transform the position from one coordinate system to another:

$$\begin{pmatrix} X \\ Y \\ Z \end{pmatrix} = [T] \begin{pmatrix} x \\ y \\ z \end{pmatrix} \quad (5.6)$$

The transformation matrix T in this form was implemented in the code of application to ease recalculation of the resource position.

### 5.6.3. Displaying relations

`InteractionModel` is responsible for determining how to display relations. In the implemented system, there are two types of relations to display - one is based on common references in resources, and the other on references to time.

### Highlight

If the user clicks on resource to display relations, the related resources are highlighted. When user clicks on the resource once again, the highlight is turned of.

Click event sets a flag on the device that starts sending a message to the server that certain device was clicked. On receiving this message, `InteractionModel` checks relations of this resource and for each related one adds a command to the array sent to the proper device, i.e. the one on which the considered resource is currently visible. The server keeps sending the command as long as it receives the message about clicking from the device.

### Timeline

For temporal relations, it was decided to show them in a form of timeline. It means that consecutive events (from those containing temporal information) are connected with lines, within or across the devices. For each two consecutive events, the coordinates of line are determined.

If the server finds that both resources are visible on the same device, it sends a message to draw line between them. The issue is when the line should go across devices.

Because the coordinates of both resources in global coordinate system are known, we can determine the coordinates of begin and end of the line in this coordinate system. However, the device has to receive message telling how it should draw line on its screen.

Let us consider resources A visible on device 1 and B visible on device 2. We should draw a line on device 1 from resource A pointing towards resource B, and consequently a line on device 2 from resource B pointing towards resource A.

To do this, we take global coordinates of resource B and calculate its position in the local coordinate system of device 1. From mathematical point of view, we perform an inverse operation to this calculating global position of resource. Then, we send message to device 1 to draw line between resource A and the calculated coordinates. The way of implementation of line drawing in Android application makes it possible to pass coordinates exceeding the boundaries of screen - it will draw the visible part of the line. Similarly, we draw the other part of the line on device 2.

Let us denote the global coordinates of a resource A as $G_A$. We obtained these coordinates by multiplying the local position on device 1 by transformation matrix of device 1, calculated basing on the position and Euler angles of this device:

$$[G_A] = [T_1][L_{A1}] \qquad (5.7)$$

We can calculate the local position of resource A on the screen of device 1 by inversing this operation:

$$[L_{A1}] = [T_1]^{-1}[G_A] \qquad (5.8)$$

However, if we take the transformation matrix calculated for another device, we will obtain the position of the resource in the coordinate system of this device, even though it is far beyond the screen boundaries.

$$[L_{A2}] = [T_2]^{-1}[G_A] \qquad (5.9)$$

Thanks to this operation, we obtain coordinates, which we can use to draw the line on the device's screen.

### 5.6.4. Passing resource between devices

As the system should enable intuitive way of passing resources between devices, it was important to find an effective method of determining where the thrown resource should appear.

If the user swipes the resource on the screen of device with velocity exceeding an empirically determined value, the resource disappears and a message is sent to server that this resource has been thrown. This message contains also the values of x and y velocity of user's finger on screen at the moment of throwing.

Therefore the server receives the vector of velocity of the resource. This vector is then rotated by the device's angles to give the proper direction in the global coordinate system.

This vector is originated in the centre of the device it comes from. Then, the system calculates the angle between the vector and X axis. In the next step, we temporarily translate the coordinate system to the centre of the device from which the resource has been thrown, and rotate it by the angle of throw vector, so that X axis is the direction of movement of resource. Then, angles for each device are calculated in this coordinate system. For this, java.Math `atan2` function is used, which is an important factor there, as is returns angle value from -pi to pi. If there is no device with absolute value of angle smaller than pi/2, the resource remains visible on the original device. Otherwise, the device with smaller absolute value of angle is chosen and a message is sent to make the resource visible.

## 5.7. Final prototype implementation

The final system developed for the user study was called Ramparts and launched on three tablets. Google Nexus 9 were used as the devices, as they are running Android OS without any modifications. Wireless communication was established using a Linksys router. One PC computer running Windows operating system was connected to the camera system with Ethernet and to the Ramparts system by Wi-Fi. Ramparts server was running on another computer, connected to the same Wireless Local Area Network (WLAN).

After start of recording in QTM, 6DOF data about each device could be requested from the machine with with Qualisys software. Then, the server and applications on tablets were launched and they started communication.

With the system running in this configuration, three basic interactions were possible: manipulating post-its visible on the tablets both on the screen and across the devices. When the user thouched the post-it, related post-its were highlighted on all devices. When the user long touched on the post-it with temporal information. All related post-its were highlighted and connected with lines in order of happening. The relations between resources were stored on the server.

# 6. Evaluation

As the result of work on this project, a functional system was developed. This system was then evaluated as a part of user study conducted in a research project. The purpose of the user study was to investigate whether and how the system design influences the task completion process, compared to the systems that are currently available. The process of study design and conducting experiments finds a valuable support in literature [16,45].

## 6.1. Study apparatus

The system was set up in an isolated environment. For the purposes of user study, two versions of the system were prepared, first for user learning and the other for the proper study. Apart from that, a setup for recording the interactions on the table was prepared. In Eclipse IDE, run settings were configured to store the output, containing among others all commands sent to devices, to a text file for further statistical analysis.

The baseline conditions were:

- tabletop interactive system, which does not offer modality, has reduced tangibility over Ramparts and does not have the innovative feature of relation visualization [Figure 17],
- paper, which does not offer any form of data preprocessing and interactive visualization [Figure 18].

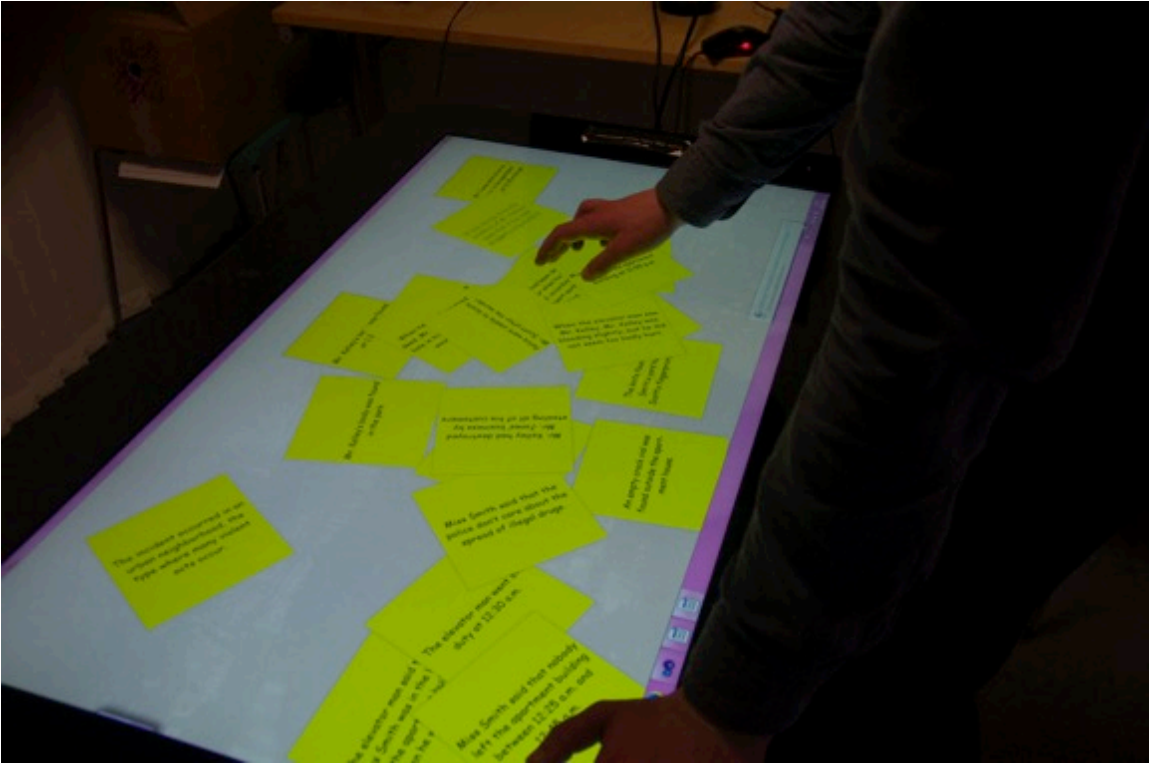

*Figure 17 Interactive tabletop baseline apparatus*

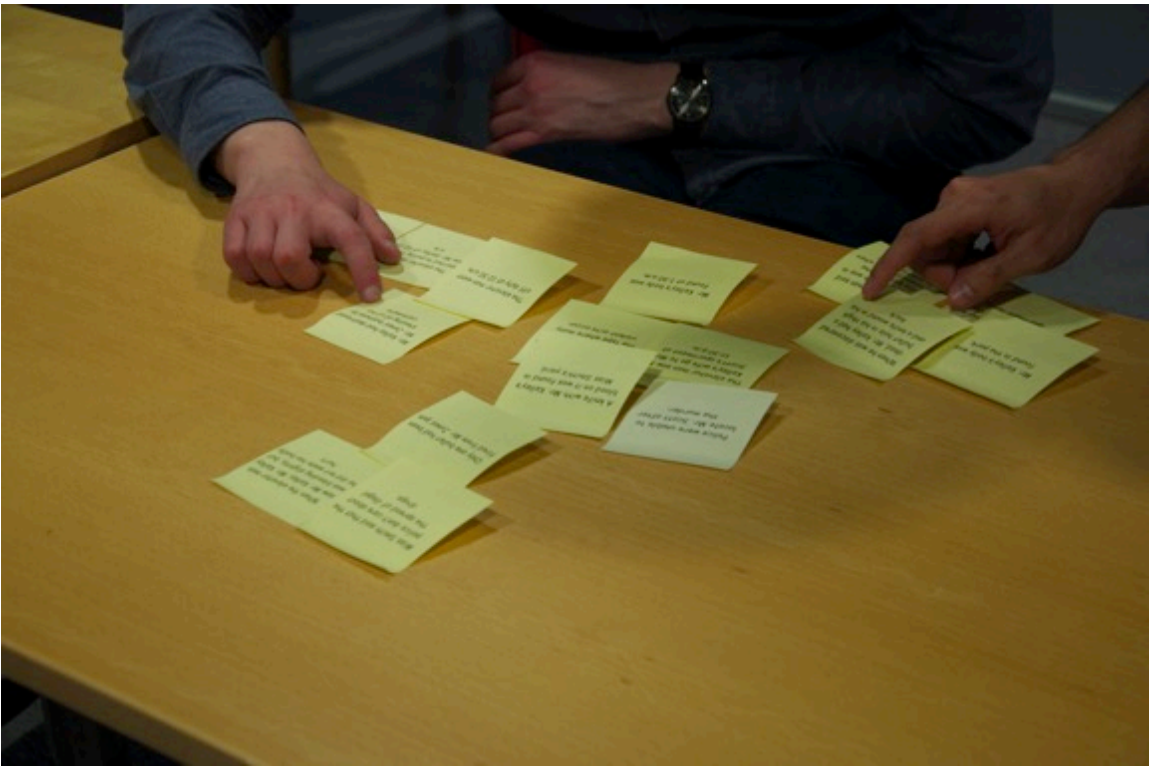

*Figure 18 Paper post-its baseline apparatus*

### 6.2. Study design

The research hypothesis of the study was that a system offering tangibility, modality, spatial awareness and possiblility to visualise data relations affects collaborative sensemaking: it enhances performance in task completion and reduces cognitive demand.

The null hypothesis was that it does not affect it, so in all the conditions performance and cognitive demand are similar.

The task in the study was to solve a crime mystery, in which we have 31 clues containing information about events, people that took part in the events, and sometimes time of event. The questions that were to be answered included: the person of murderer, the time of murder, the place and the motive. The study task was based on a criminal mystery from a book prepared for teaching collaboration skills [31].

The desired participant profile for the study was defined as follows: aged from undergraduate students to people before retirement, with basic, but well established knowledge about usage of mobile devices and with no professional experience in data analysis. The participants were given surprise gifts as a form of remuneration.

The study consisted of three main parts. In the first part, the users were introduced to the structure of the study and to the system. Then, they were shown all features offered by the system and left alone to learn how to use them. The second part was the proper task: the users were introduced to the situation and given answer sheet, then asked to start the task. The third and last part of the study consisted of questionnaires concerning the cognitive demand and mutual collaboration, and of semi-structured interview investigating their feelings about the proposed setup.

## 6.3. Study results

In the study, 18 participants in 9 pairs (aged 24-61, µ=33, $\tilde{x} = 37,2$, 10 males, 8 females) used the system to solve the criminal task. Each pair managed to come to proper answers in time shorter than 30 minutes. Each pair of participants was introduced to the functionalities of the system and used each of them during task completion. After conducting the study with baseline systems, the results will be published in a scientific paper.

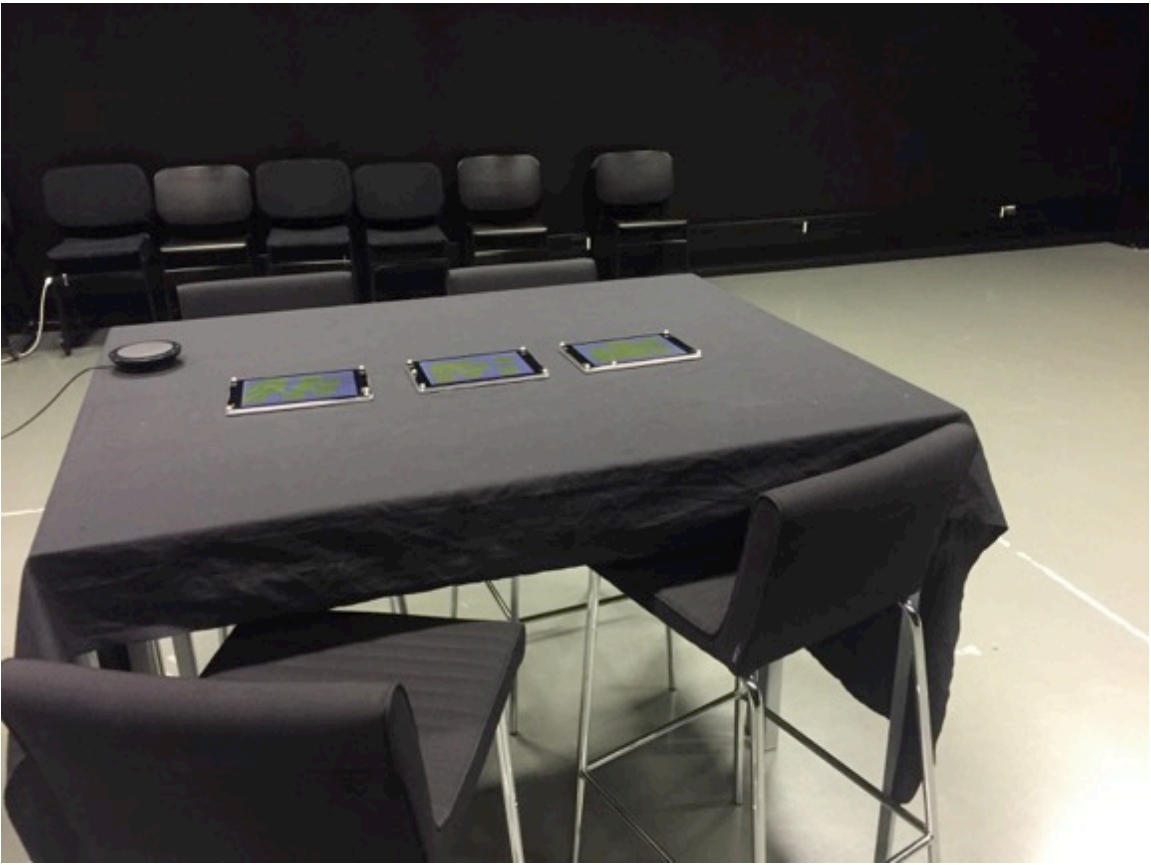

*Figure 19 User study setup*

From the interviews with participants, some preliminary conclusions may be drawn. Many participants liked the spatial features of the system. The functionality used most often was highlighting of related post-its. Users also tended to move the resources between the devices. They also gave a considerable input regarding further development of the system.

# 7. Conclusions

In the course of the project, an interactive system based on multiple spatially-aware mobile devices was designed and implemented. The direction of the work was determined by literature research in the related work in the field of Human-Computer Interaction. This research revealed and unexplored niche, which I, working as a part of a research team, decided to investigate.

The design of the system was also preceded by examination of the related work on design considerations for cross-device interactions. The design was furthermore based on testing with low fidelity prototypes and preliminary user study.

The final design specified a system for enhancing collaborative sensemaking on multiple mobile devices, with employing their spatial awareness. Such systems are of considerable value for tackling with task that require human expertise, such as thematic analysis and decision making.

The implemented system consisted of a server managing the devices and receiving positional information, developed in Java programming language. The mobile application was developed for Android OS. The final system supported three main interactions: moving the elements displayed on screen to another device, highlighting related pieces of information and displaying a visual timeline across the devices.

The development of the system involved proper handling of the aforementioned interactions, which required designing decision making algorithms based on mathematical calculations of the position and direction of interactions happening in the scope of the system.

The system was evaluated during a user study conducted in the course of a research project. The detailed results of the study, including comparison with described baseline systems, will be published in a scientific paper.

## Further work

Further work on this subject may be divided into two main areas. First is implementing new functionalities to the already developed system. Many suggestions about possible solutions were given by the participants of the user study.

The second area of further development is extending the system in the means of hardware. This may include evaluating the interactions on a larger number of devices. Also, further work may involve engaging other types of mobile devices, such as wearables, to the interactive system presented in this thesis.

# 9. Index of tables and figures

## Index of tables

## Index of figures